\documentclass[12pt]{article}
\usepackage{amssymb,amsmath,amsthm,amsfonts,amscd}
 \usepackage{graphicx}
\newcommand\be{\begin{equation}}
\newcommand\ee{\end{equation}}
\newcommand\bea{\begin{eqnarray}}
\newcommand\eea{\end{eqnarray}}

\newcommand{\fatalpha}{{\bf \alpha \kern -0.44em \alpha}}
\newcommand{\fatsigma}{{\bf \sigma \kern -0.54em \sigma}}
\newcommand{\tpchi}{{\bf D \kern -0.35em D}}
\newcommand{\llambda}{{\bf \lambda \kern -0.45em \lambda}}

\bibliography{plain}
\title{\bf Exact dynamics of one-qubit system in  layered  environment  } \vspace{20mm}
\author{ M.Mahdian$^{a}$
 \thanks{E-mail:Mahdian@tabrizu.ac.ir},
 ,H. Mehrabpour$^{a}$ ,
 \\
$^a${\small Department of Theoretical Physics and Astrophysics,
University of Tabriz, Tabriz 51664, Iran.}  } \pagebreak

\vspace{20mm}

\begin{document}
\maketitle \vspace{15mm}
\newpage
\begin{abstract}
We investigate the exact evolution of the reduced dynamics of a one
qubit system as central spin coupled to a femionic layered
environment with unlimited number of layers. Also, we study the
decoherence induced on central spin by analysis solution is obtained
in the limit $N\rightarrow\infty$ of an infinite number of bath
spins. Finally, the Nakajima-Zwanzig (NZ) and the
time-convolutionless (TCL) projection operator techniques to second
order are derived.


\end{abstract}
\section{Introduction}  
In quantum mechanics, quantum information is physical information
that is held in the state of a quantum system\cite{MANielsen}.
Quantum information theory focuses on the amount of accessible
information\cite{DLoss}, it can be regarded as the theory for
quantitative evaluation of the process of extracting
information\cite{BEKane,DDAwschalom}. Every quantum system
encountered in the real world is an open quantum system and the
theory of open quantum systems describes how a system of interest is
influenced by the interaction with its environment. This interaction
often leads to a loss of the quantum features of physical states and
has a great impact on the dynamical behavior of the open system due
to the non-unitary characteristic of the time evolution, although
much care is taken experimentally to eliminate the unwanted
influence of external interactions, there remains, if ever so
slight, a coupling between the system of interest and the external
world\cite{FPetruccione,ECGSudarshan}. One kind of open quantum
systems study to describe the information extraction process in
quantum information are the spin star systems, where a central spin
-$\frac{1}{2}$ particle couples to a spin bath of N
spin-$\frac{1}{2}$ particles and they have attracted a vast amount
attention in the quantum community \cite{AHutton}-\cite{VSemin}
because they are of significance and of interest due to their high
symmetry, strong non-Markovian behavior and also as one of the best
candidates of the spin-qubit quantum
computation\cite{HPBeruer}-\cite{DRossini}. This is even more
relevant when environmental influences of a non-Markovian nature,
such as those due to memory-keeping and feedback-inducing
system-environment mechanisms, are considered\cite{FPetruccione}.

The spin star configuration can also describe decoherence model
\cite{CMDawson} because the coupling of an open quantum system with
its environment causes correlations between the states of the system
and the bath\cite{FPetruccione}. the correlations exchange the
information between the open quantum system and its environment and
the environment-induced, dynamic destruction of quantum coherence is
called decoherence\cite{WHZurek}. In the language of state and
density matrix, the superposition of the open quantum system's
states is destroyed after tracing over the environmental degrees of
freedom and the system's reduced density matrix turns into a
statistical mixture.

Motivated by this consideration, in this paper, we consider layered
environment with a spin at the center of layers to study a
generalized spin star system which can be solved exactly. It must be
noted that in the model, degeneracy for coupling coefficients are
considered.

The paper is organized as follows. In Sec.(2), we introduce the
model investigated, a spin star model involving a Heisenberg XX
coupling in Sec.(2.1), and determine the exact time evolution of the
central spin in Sec.(2.2). Therefore, if we equalize all coupling
coefficients with together, we obtain the result of
Ref.\cite{HPBeruer}. In Sec.(3) we assume two different layers of
environment and compute our model with this assumption. Furthermore,
we analyze the limit of an infinite number of bath spins, discuss
the behavior of the von Neumann entropy of the central spin, and
demonstrate that the model exhibits complete relaxation and partial
decoherence. The non-Markovian approximation techniques are
discussed in Sec.(4). In this section the dynamic equations found in
the second order of the coupling are introduced. It is also
demonstrated that the prominent Born-Markov approximation is not
applicable to the spin star model. Of course, the Born-Markov
approximation is second order Nakajima-Zwanzig.
\section{Exact Dynamics}

\subsection{The Model}
We consider a spin star configuration which consists of N+1
localized spin-$\frac{1}{2}$ particles. One of the spins is located
at the center of the star, while the others are on concentric
circles with different radii surrounding the central spin, layer by
layer the difference in radius is because the coupling coefficients
between layers spins and the central spin are taken differently. It
must be noted that in this model degeneracy for coupling
coefficients is considered, because naturally some of spins are in
relation with the central spin by a constant coupling coefficient
which are located in one layer. By considering such model, the most
general model of single-qubit spin-Star for fermionic particles with
fixed fermionic environment is made, Fig(1). This model explains how
particles of bath with different coupling coefficients can be used
to control time of decoherence. Because degeneracy coefficient
specifies the number of the particle layer, decoherence-time by each
layer can be controlled with different degeneracy coefficient and we
should not lose the effect of degenerate factor. However, we
consider our model by the below description and explanation. The
central spin $\sigma$ interacts with the bath spins $\sigma^{(j)}$
via a Heisenberg XX interaction \cite{EHLieb}represented through the
Hamiltonian
\begin{equation}\label{Hamiltonian}
H=2(\sigma_{+}\Xi_{-}+\sigma_{-}\Xi_{+}),
\end{equation}
where $\Xi_{\pm}$ are denoted as follows
\begin{equation}\label{j positive}
\Xi_{+}=\sum_{\mu=1}^{n}\alpha_{\mu}J^{\mu}_{+},
\end{equation}
\begin{figure}[htb]
\begin{center}
\includegraphics{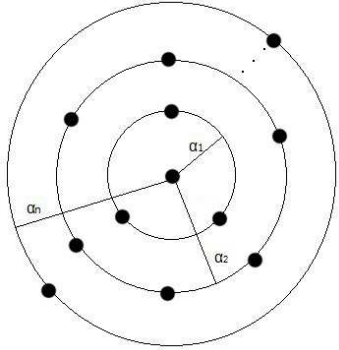}
      \vspace{4cm}
\caption{The figure depicts the general layered environment with one
spin in the center of layers.} \label{fig1}
\end{center}
\end{figure}
\begin{equation}\label{j negative}
\Xi_{-}=\sum_{\mu=1}^{n}\alpha_{\mu}J^{\mu}_{-}.
\end{equation}
Here, $\alpha$ coefficients specify interaction between system and
environment and is dependent from distance. Also we have
\begin{equation}\label{Hamiltonian}
J^{\mu}_{\pm}\equiv\sum_{j=1}^{N_{\mu}}\sigma^{j}_{\pm},
\end{equation}
here, $\mu=1,2,\ldots,n$ for n different layers of bath. Also, we
have
$$
\sigma^{j}_{\pm}\equiv\frac{1}{2}(\sigma^{j}_{1}\pm
i\sigma^{j}_{2}),
$$
that represents the raising and lowering operators of the jth bath
spin. The Heisenberg XX coupling has been found to be an effective
Hamiltonian for the interaction of some quantum dot systems
\cite{AImamoglu}. Equation(1) describes a very simple time
independent interaction with equal coupling strength $\alpha_{1}$
for $N_{1}$ of first bath spin and $\alpha_{2}$ for $N_{2}$ of
second bath spin to $\alpha_{n}$ for $N_{n}$ of nth bath spin. It is
invariant under rotation around the $\textit{z}-axis$. The operator
$J\equiv\frac{1}{2}\sum_{\mu=1}^{N}\sigma^{\mu}$ represents the
total spin angular momentum of the bath (units are chosen such that
$\hbar=1$). Therefore the central spin thus couples to the
collective bath angular momentum.

We introduce an Orthonormal basis in the bath Hilbert space $ H_{B}
$ consisting of states $|j_{\mu},m_{\mu},x\rangle$ where $\mu$ is 1
to n. These states are defined as eigenstates of $\textbf{J}_{3}$
(eigenvalue m) and of $\textbf{J}^{2}$ (eigenvalue j(j+1)). The
index $x$ labels the different eigenstates in the eigenspace
$\textbf{{\textsl{K}}}_{j,m}$ belonging to a given pair (j,m) of
quantum numbers. As usual, $j_{\mu}\leq\frac{N_{\mu}}{2}$ and
$-j_{\mu}\leq m \leq j_{\mu}$ where $\mu$ is 1 to n. The dimension
of $\textbf{\textsl{K}}_{j,m}$ is given by the expression
\cite{Hutton,JWesenberg}
\begin{equation}\label{degeneracy}
\Upsilon(j_{\mu},N_{\mu})=\left(
        \begin{array}{c}
N_{\mu}\\
\frac{N_{\mu}}{2}-j_{\mu}
        \end{array}
      \right)-\left(
        \begin{array}{c}
N_{\mu}\\
\frac{N_{\mu}}{2}-j_{\mu}-1
        \end{array}
      \right).
 \end{equation}
We assume that the initial state of the composite system be a
product state. That is
\begin{equation}\label{composite system}
\rho(0)=\rho_{S}(0)\otimes \rho_{B}(0).
\end{equation}
We can calculate the reduced density matrix of the quantum system in
the following expression.
\begin{equation}\label{quantum system}
\rho_{S}(t)=tr_{B}(U \rho(0)U^{\dag}).
\end{equation}
The above equation is obtained by doing partial-trace on bath and
also U is the unitary operator which is defined as follows
$$
U=\exp(-iHt).
$$
The reduced density matrix is completely determined in terms of the
Bloch vector
\begin{equation}\label{Bloch vector}
\chi(t)=\left(
        \begin{array}{c}
\omega_{1}(t)\\
\omega_{2}(t)\\
\omega_{3}(t)\\
        \end{array}
      \right)\equiv Tr(\sigma\rho_{S}(t)),
 \end{equation}
through the relationship
\begin{equation}\label{system at t}
\rho_{S}(t)=\frac{1}{2}\left(
        \begin{array}{cc}
1+\omega_{3}(t)&\omega_{1}(t)-i\omega_{2}(t) \\
\omega_{1}(t)-i\omega_{2}(t)&1-\omega_{3}(t)\\
        \end{array}
        \right),
\end{equation}
We note that the length $q(t)\equiv|\chi(t)|$ of the Bloch vector is
equal to 1 iff $\rho_{S}(t)$ describes a pure state, and the von
Neumann entropy $\emph{S}$ of the central spin can be expressed as a
function of the length q(t) of the Bloch vector:
\begin{equation}\label{Neumann entropy}
\emph{S}\equiv
Tr[-\rho_{S}\ln\rho_{S}]=\ln2-\frac{1}{2}(1-q)+\frac{1}{2}(1+q)\ln(1+q).
\end{equation}
The initial state of the reduced system at t=0 is taken to be an
arbitrary (possibly mixed) state
\begin{equation}\label{system at zero}
\rho_{S}(0)=\left(
        \begin{array}{cc}
\frac{1+\omega_{3}(0)}{2}&\omega_{-}(0) \\
\omega_{+}(0)&\frac{1-\omega_{3}(0)}{2}\\
        \end{array}
        \right),
\end{equation}
while the spin bath is assumed to be in an unpolarized infinite
temperature state:
\begin{equation}\label{bath}
\rho_{B}(0)=2^{-N}I_{B}.
\end{equation}
Here, $I_{B}$ denotes the unit matrix in $ H_{B} $ and N is
$N_{1}+N_{2}+\ldots+N_{n}$, and we have defined the $\omega_{\pm}$
as linear combinations of the components $\omega_{1,2}$ of the Bloch
vector
\begin{equation}\label{Bloch vector+,-}
\omega_{\pm}=\frac{\omega_{1}\pm i\omega_{2}}{2}.
\end{equation}
\subsection{Reduced System Dynamics}

In this section, we will derive the exact dynamics of the reduced
density matrix $\rho_{S}(t)$ for our given model. We obtain the
evolution of central spin with n different coupling coefficients
that should be used from Eq.(7) until the solution model is exact.
This yeilds
\begin{equation}\label{density matrix}
\rho_{S}(t)=tr_{B}\{i^{l}\sum_{l=0}^{\infty}\sum_{n=0}^{l}\frac{t^{l}}{l!}(-1)^{n}\left(
        \begin{array}{c}
l\\
n\\
        \end{array}
      \right)H^{n}(\rho_{S}(0)\otimes \frac{I_{B}}{2^{N}})H^{l-n}\}.
      \end{equation}
It can easily be verified that
\begin{equation}\label{hamiltonian even}
H^{2k}=4^{k}[\sigma_{+}\sigma_{-}(\Xi_{-}\Xi_{+})^{k}+\sigma_{-}\sigma_{+}(\Xi_{+}\Xi_{-})^{k}],
\end{equation}
and
\begin{equation}\label{hamiltonian odd}
H^{2k+1}=2\cdot4^{k}[\sigma_{+}\Xi_{-}+\sigma_{-}\Xi_{+}].
\end{equation}
We note that such simple expressions are obtained since a term
$\sigma_{3}J_{3}$ is missing in the interaction Hamiltonian. We
substitute the last two equations into Lindblad
equation\cite{HPBeruer} as follows
\begin{equation}\label{density matrix}
L^{l}\rho=i^{l}\sum_{n=0}^{l}(-1)^{n}\left(
        \begin{array}{c}
l\\
n\\
        \end{array}
      \right)H^{n}\rho H^{l-n},
      \end{equation}
to get the formulas
\begin{equation}\label{zero term}
tr_{B}\{\ L^{2l+1}\rho_{S}(0)\otimes \ 2^{-N}I_{B}\}=0,
\end{equation}
and
\begin{equation}
tr_{B}\{\ L^{2l}\rho_{S}(0)\otimes \ 2^{-N}I_{B}\}=
\sum_{l=0}^{\infty}(-4)^{l}\sum_{n=0}^{2l}\frac{t^{2l}}{(2l)!}\left(
        \begin{array}{c}
2l\\
2n\\
\end{array}
\right)[(\frac{1+\omega_{3}\sigma_{3}}{2})\Omega_{l}+(\omega_{+}\sigma_{-}+\omega_{-}\sigma_{+})\Gamma^{l-n}_{n}],
\end{equation}

which hold for all l=1,2,\ldots. Here,we have introduced the bath
correlation functions
\begin{equation}\label{omega}
\Omega_{l}\equiv\frac{1}{2^{N}}tr_{B}\{(\Xi_{+}\Xi_{-})^{l}\},
\end{equation}
\begin{equation}\label{gamma}
\Gamma^{l-n}_{n}\equiv\frac{1}{2^{N}}tr_{B}\{(\Xi_{+}\Xi_{-})^{l-n}(\Xi_{-}\Xi_{+})^{n}\},
\end{equation}
where we have product $\Xi_{\pm}\Xi_{\mp}$ as follows
\begin{equation}
\Xi_{\pm}\Xi_{\mp}=(\sum_{\mu=1}^{n}\alpha_{\mu}J^{\mu}_{\pm})(\sum_{\mu=1}^{n}\alpha_{\mu}J^{\mu}_{\mp}).
\end{equation}
Of course, Eq.(18) is zero because it is
$$
\langle j,m|J_{\pm}|j,m\rangle=0.
$$

We will come back to these correlation functions when we discuss
approximation techniques in Sec.(4).

Using the formulas (18) and (19) in Eq.(14) we can express the
components of the Bloch vector as follows,
\begin{equation}
\omega_{\pm}(t)=f_{\pm}(t)\omega_{\pm}(0),
\end{equation}
\begin{equation}
\omega_{3}(t)=f_{z}(t)\omega_{3}(0),
\end{equation}
where we have introduced the functions

\begin{equation}
f_{\pm}(t)\equiv
tr_{B}\{\cos[2th_{1}(\alpha_{1},\ldots,\alpha_{n})]\cos[2th_{2}(\alpha_{1},\ldots,\alpha_{n})]
\otimes 2^{-N}I_{B}\},
\end{equation}

and

\begin{equation}
f_{z}(t)\equiv
tr_{B}\{\cos[2th_{1}(\alpha_{1},\ldots,\alpha_{n})]\otimes
2^{-N}I_{B}\},
\end{equation}
where $h_{1}(\alpha_{1},\alpha_{2},\ldots,\alpha_{n})$ and
$h_{2}(\alpha_{1},\alpha_{2},\ldots,\alpha_{n})$ are
\begin{equation}
h_{1}(\alpha_{1},\alpha_{2},\ldots,\alpha_{n})=\sqrt{\sum_{\mu=1}^{n}\alpha^{2}_{\mu}J^{\mu}_{+}J^{\mu}_{-}},
\end{equation}
\begin{equation}
h_{2}(\alpha_{1},\alpha_{2},\ldots,\alpha_{n})=\sqrt{\sum_{\mu}^{n}\alpha^{2}_{\mu}J^{\mu}_{-}J^{\mu}_{+}}.
\end{equation}
Calculating the traces over the spin bath in the eigenbasis of
$J_{3}$ and $\textbf{J}^{2}$ using
\begin{equation}\label{JJ}
J_{\pm}J_{\mp}|j,m,x\rangle=(j\pm m)(j\mp m+1)|j,m,x\rangle.
\end{equation}
We find
\begin{equation}\label{f12 for n}
f_{\pm}(t)=[\prod_{i=1}^{n}\sum_{j_{i},m_{i}}\Upsilon(N_{i},j_{i})]\frac{\cos(4t\sqrt{\zeta})\cos(4t\sqrt{\eta})}{2^{\sum_{i=1}^{n}}N_{i}},
\end{equation}
and
\begin{equation}\label{f3 for n}
f_{z}(t)=[\prod_{i=1}^{n}\sum_{j_{i},m_{i}}\Upsilon(N_{i},j_{i})]\frac{\cos(4t\sqrt{\zeta})}{2^{\sum_{i=1}^{n}}N_{i}},
\end{equation}
here,$\zeta$ and $\eta$ denoted are as
$$
\zeta=\sum_{i=1}^{n}h_{i}^{+}\alpha_{i}^{2},
$$
and
$$
\eta=\sum_{i=1}^{n}h_{i}^{-}\alpha_{i}^{2},
$$
and also, we have
$$
h_{i}^{+}=h(j_{i},m_{i}); h_{i}^{-}=h(j_{i},-m_{i}),
$$
where we have introduced the quantity $h(j,\pm m)=(j\pm m)(j\mp
m+1)$.

Thus we have determined the exact dynamics of the reduced system:
The density matrix $\rho_{S}(t)$ of the central spin is given
through the components of the Bloch vector which are provided by the
relations (23),(24) and (30),(31). We note that the dynamics can be
expressed completely through only two real-valued function
$f_{\pm}(t)$ and $f_{z}(t)$. This fact is connected to the
rotational symmetry of the system. Also from the overall role of
coupling coefficients and degeneracy coefficients in the relations
(30) and (31), it will be shown that the study of control over
decoherence is on coupling coefficients and degeneracy coefficients
that you'll see Sec.(3).


\section{Example}

In this example, we consider the central spin with two different
bath by two different coupling coefficients. This means we consider
two layers with different radii. The choice of different coupling
coefficients certainly will affect on degeneracy coefficient because
the behavior of particles in each layer is different from other
layers and this difference is to express the considering by
different coupling coefficients and different degeneracy
coefficients. So,we consider our hamiltonian as
\begin{equation}
H=2(\sigma_{+}\Xi_{-}+\sigma_{-}\Xi_{+}),
\end{equation}
where we define $\Xi$ as
$$
\Xi_{+}=\sum_{\mu=1}^{2}\alpha_{\mu}J^{\mu}_{+},
$$
$$
\Xi_{-}=\sum_{\mu=1}^{2}\alpha_{\mu}J^{\mu}_{-}.
$$
According to the introduced model, we can express two real-valued
functions $f_{\pm}(t)$ and $f_{z}(t)$ as
\begin{equation}\label{f12}
f_{\pm}(t)=\sum_{j_{1},m_{1}}\sum_{j_{2},m_{2}}\Upsilon(N_{1},j_{1})\Upsilon(N_{2},j_{2})\frac{\cos(2t\sqrt{\beta})\cos(2t\sqrt{\gamma})}{2^{N_{1}+N_{2}}},
\end{equation}
and
\begin{equation}\label{f3}
f_{z}(t)=\sum_{j_{1},m_{1}}\sum_{j_{2},m_{2}}\Upsilon(N_{1},j_{1})\Upsilon(N_{2},j_{2})\frac{\cos(4t\sqrt{\beta})}{2^{N_{1}+N_{2}}},
\end{equation}
where $\beta$ and $\gamma$ are denoted as
$$
\beta=\alpha^{2}_{1}h(j_{1},m_{1})+\alpha^{2}_{2}h(j_{2},m_{2}),
$$
and
$$
\gamma=\alpha^{2}_{1}h(j_{1},-m_{1})+\alpha^{2}_{2}h(j_{2},-m_{2}).
$$
The explicit solution constructed in the previous section takes on a
relatively simple form in the limit $N\rightarrow\infty$ of an
infinite number of bath spins \cite{HPBeruer}. Because N is
$N_{1}+N_{2}$, we should discuss about $N_{1}\rightarrow\infty$ or
$N_{2}\rightarrow\infty$ or both. in Ref.\cite{HPBeruer} for large
$N$, the corresponding correlation function is obtained. But here by
considering two layers, we obtain the states of
$N_{1}\rightarrow\infty$ and $N_{2}\rightarrow\infty$ for
corresponding correlation functions as follows:
\begin{equation}
\Omega_{l}\approx\sum_{k=0}^{l}\left(
        \begin{array}{c}
l\\
k\\
\end{array}
\right)(l-k)!k!\frac{\alpha_{1}^{2(l-k)}\alpha_{2}^{2k}}{2^{l}},
\end{equation}
\begin{equation}
\Gamma_{n}^{l-n}\approx\sum_{k=0}^{n}\sum_{\acute{k}=0}^{l-n}\left(
        \begin{array}{c}
n\\
k\\
\end{array}\right)\left(
        \begin{array}{c}
l-n\\
\acute{k}\\
\end{array}\right)\frac{(n-k)!(l-n-\acute{k})!k!\acute{k}!\alpha_{1}^{2(n-k)}\alpha_{2}^{l-n-\acute{k}}}{2^{l}}.
\end{equation}

Of course, we assume a non-trivial finite limit
$N_{i}\rightarrow\infty$, therefore we rescale the coupling constant
as\cite{HPBeruer},
\begin{equation}
\alpha_{i}\rightarrow\frac{\alpha_{i}}{\sqrt{N_{i}}}.
\end{equation}
Using this approximation in Eq.(19), we can rewrite $f_{\pm}$ and
$f_{z}$ as
\begin{equation}
f_{\pm}(t)=\sum_{l=0}^{\infty}\sum_{n=0}^{2l}\sum_{k=0}^{n}\sum_{\acute{k}=0}^{l-n}\frac{(-2)^{l}t^{2l}}{(2l)!}\left(
        \begin{array}{c}
2l\\
2n\\
\end{array}
\right)\left(
        \begin{array}{c}
n\\
k\\
\end{array}
\right)\left(
        \begin{array}{c}
l-n\\
\acute{k}\\
\end{array}
\right)k!\acute{k}!(n-k)!(l-n-\acute{k})!\alpha_{1}^{2(l-n-\acute{k})}\alpha_{2}^{2(n-k)},
\end{equation}
and
\begin{equation}
f_{z}(t)=\sum_{l=0}^{\infty}\sum_{k=0}^{l}\frac{(-8)^{l}t^{2l}}{(2l)!}\left(
        \begin{array}{c}
2l\\
2n\\
\end{array}
\right)(l-k)!k!\alpha_{1}^{2(l-k)}\alpha_{2}^{2k}.
\end{equation}
The up state is a state with consideration to
$N_{1}\rightarrow\infty$ and $N_{2}\rightarrow\infty$, which is for
the most general level cases non-trivial for an infinite limit. But
there might be a layer which has a finite number of particle and
another one with an infinite number of particle (of course here, it
is understandable that infinite means the order of Avogadro's number
because this size of particles in the scale of our study, accounts
for infinity). We assume that $N_{2}$ is limited and $N_{1}$ is
unlimited. Of course in computing because of the symmetry of the
system, there isn't any difference is taking $N_{1}$ limited and
$N_{2}$ unlimited with the previous case. So, for
$N_{1}\rightarrow\infty$ we can obtain as
$$f_{\pm}(t)=\frac{1}{2^{N_{2}}}\sum_{j_{2},m_{2}}\sum_{n=0}^{\infty}\sum_{k=0}^{\infty}\Upsilon(N_{2},j_{2})\cos(2t\alpha_{2}\sqrt{h(j_{2},m_{2})})
\cos(2t\alpha_{2}\sqrt{h(j_{2},-m_{2})})$$
\begin{equation}
\times\frac{(-2)^{n}t^{2n}}{(2n)!}n!\frac{(-2)^{k}t^{2k}}{(2k)!}k!\alpha_{1}^{2(n+k)},
\end{equation}
and
\begin{equation}
f_{z}(t)=\frac{1}{2^{N_{2}}}\sum_{j_{2},m_{2}}\Upsilon(N_{2},j_{2})\cos(4t\alpha_{2}\sqrt{h(j_{2},m_{2})})
\{1-2\sqrt{2}t\alpha_{1}DF[\sqrt{2}t\alpha_{1}]\}.
\end{equation}
Note that Dawson Function is closely related to the error function
erf, as
$$
DF(x)=\frac{\sqrt{\pi}}{2}\exp(-x^{2})erfi(x),
$$
where erfi is the imaginary error function, $erfi(x)=-ierf(ix)$.

\section{approximation techniques}

In this section we will apply different approximation techniques to
the spin star model introduced and discussed in the previous
section. Due to the simplicity of this model we can not only
integrate exactly the reduced system dynamics, but also construct
explicitly the various master equations for the density matrix of
the central spin and analyze and compare their perturbation
expansions. In the following discussion we will stick to the Bloch
vector notation. Each of the master equations obtained can easily be
transformed into an equation involving Lindblad superoperators using
the translations rules
\begin{equation}
\rho_{S}=\frac{I+\omega_{3}\sigma_{3}}{2}+\omega_{+}\sigma_{-}+\omega_{-}\sigma_{+}.
\end{equation}
The second order approximation of the master equation for the
reduced system is usually obtained within the Born approximation
\cite{FPetruccione}. It is equivalent to the second order of the
Nakajima-Zwanzing projection operator technique. In our model the
Born approximation leads to the master equation
\begin{equation}
\dot{\rho}_{S}(t)=-\int_{0}^{t}ds
tr_{B}\{[H,[H,\rho_{S}(s)\otimes\rho_{B}(0)]]\}=-8\Omega_{1}\int_{0}^{t}ds(\omega(s)_{3}\sigma_{3}+\omega(s)_{+}\sigma_{-}+\omega(s)_{-}\sigma_{+}),
\end{equation}
where the bath correlation function is found to be
\begin{equation}
\Omega_{1}=\frac{1}{2^{N}}tr_{B}\{\Xi_{+}\Xi_{-}\}=\frac{1}{2^{N}}tr_{B}\{(\alpha_{1}J^{1}_{+}+\alpha_{2}J^{2}_{+})(\alpha_{1}J^{1}_{-}+\alpha_{2}J^{2}_{-})\}=\frac{1}{2}(\alpha_{1}^{2}N_{1}+\alpha_{2}^{2}N_{2}).
\end{equation}
It is important to notice that $\Omega_{1}$, as well as all other
bath correlation functions are independent of time. This is to be
contrasted to those situations in which the bath correlation
function decay rapidly and which allow the derivation of a Markovian
master equation. The time-independence of the correlation function
is the main reason for the non-Markovian behavior of the spin bath
model. The integro-differential Eq.(43) can easily be solved by a
Laplace transformation with the solution
\begin{equation}
f_{\pm}(t)=\frac{\omega_{\pm}(t)}{\omega_{\pm}(0)}=\cos(2t\sqrt{\delta}),
\end{equation}
\begin{equation}
f_{z}(t)=\frac{\omega_{3}(t)}{\omega_{3}(0)}\cos(2t\sqrt{2\delta}),
\end{equation}
where $\delta$ is denoted as
\begin{equation}
\delta=\alpha_{1}^{2}N_{1}+\alpha_{2}^{2}N_{2}.
\end{equation}
In many physical applications the integration of the
integro-differential equation is much more complicated and one tries
to approximate the dynamics through a master equation which is local
in time. To this end, the terms $\omega_{\pm}(t)$ and
$\omega_{3}(t)$ under the integral in Eq.(43) are replaced by
$\omega_{\pm}(t)$ and $\omega_{3}(t)$, respectively. We thus arrive
at the time-local master equation
\begin{equation}
\frac{d}{dt}\rho_{s}(t)=-4\delta\int_{0}^{t}ds(\omega_{3}(t)\sigma_{3}+\omega_{+}(t)\sigma_{-}+\omega_{-}(t)\sigma_{+})=-4\delta
t(\omega_{3}(t)\sigma_{3}+\omega_{+}(t)\sigma_{-}+\omega_{-}(t)\sigma_{+}),
\end{equation}
\begin{figure}[htb]
\begin{center}
\includegraphics{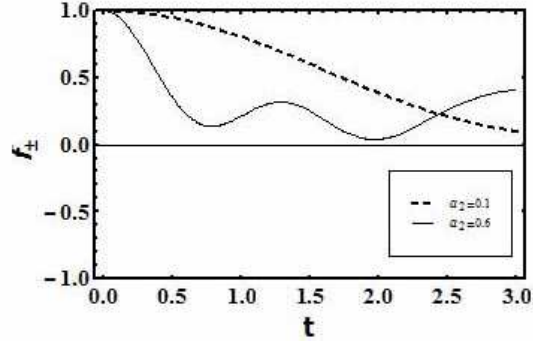}
      \vspace{4cm}
\caption{The equal spin numbers for correlation function $f_{\pm}$
[see Eqs.(33)], with $\alpha=0.1$ and $N_{i}=20$.} \label{fig4}
\end{center}
\end{figure}

\begin{figure}[htb]
\begin{center}
\includegraphics{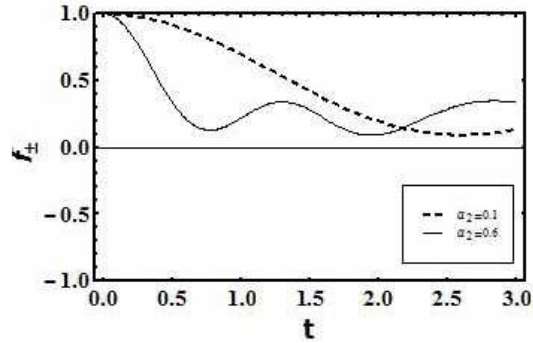}
      \vspace{4cm}
\caption{The changes of spin number of layers for correlation
function $f_{\pm}$ [see Eqs.(33)], with $\alpha=0.1$, $N_{1}=20$ and
$N_{2}=100$.} \label{fig5}
\end{center}
\end{figure}
which is sometimes referred to as Redfield equation. Also this
master equation is easily solved to give the expressions
\begin{equation}
f_{\pm}(t)=\exp(-2\delta t),
\end{equation}
\begin{equation}
f_{z}(t)=\exp(-4\delta t).
\end{equation}
The Redfield equation is equivalent to the second order of the
time-convolutionless projection operator technique. finally, In
order to obtain, a Markovian master equation, i.e. a time-local
\begin{figure}[htb]
\begin{center}
\includegraphics{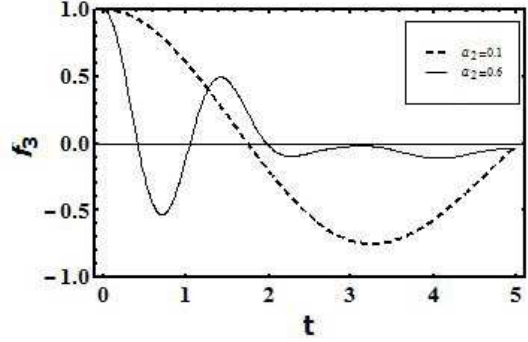}
      \vspace{4cm}
\caption{The equal spin numbers for correlation function $f_{3}$
[see Eqs.(34)],with $\alpha=0.1$ and $N_{i}=20$ .} \label{fig2}
\end{center}
\end{figure}

\begin{figure}[htb]
\begin{center}
\includegraphics{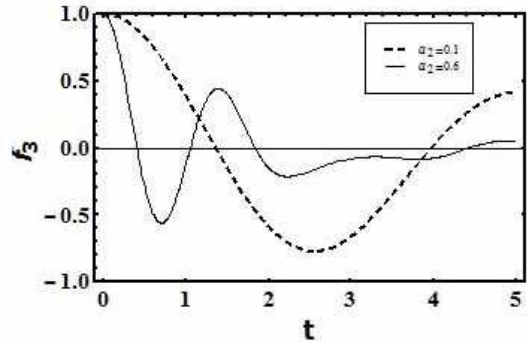}
      \vspace{4cm}
\caption{The changes of spin number of layers for correlation
function $f_{3}$ [see Eqs.(34)], with $\alpha=0.1$, $N_{1}=20$ and
$N_{2}=100$.} \label{fig3}
\end{center}
\end{figure}
equation involving a time independent generator, one pushes the
upper limit of the integral in Eq.(48) to infinity, as other studies
of the master equation. This limit leads to the Born-Markov
approximation of the reduced dynamics. In the present model,
however, it is not possible to perform this approximation because
the integrand does not vanish for large t. Thus,the Born-Markov
limit does not exist for the spin bath model investigated here and
the description of decoherence processes requires the usage of
non-Markovian methods.
\begin{figure}[htb]
\begin{center}
\includegraphics{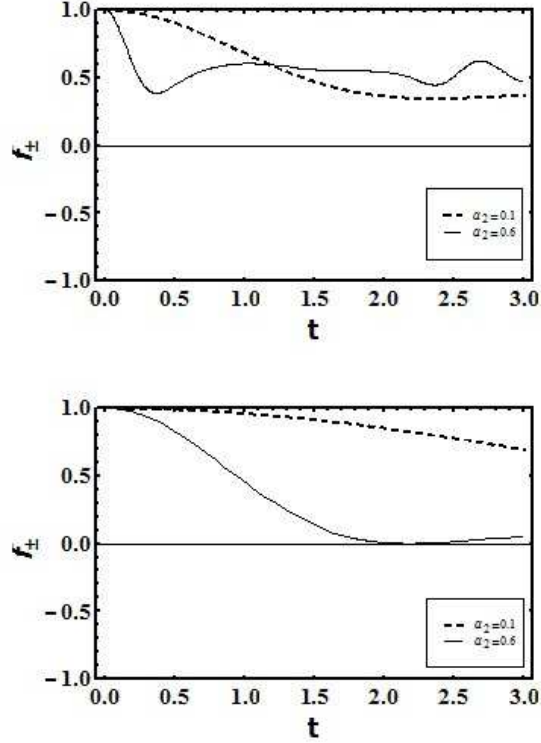}
      \vspace{10.5cm}
\caption{Comparison of limit $N\rightarrow\infty$ with limit
$N_{1}\rightarrow\infty$ or $N_{2}\rightarrow\infty$ for the changes
of correlation function $f_{\pm}$ [see Eqs.(38) and Eqs.(40)], with
$\alpha=0.1$ and $n=20$.} \label{fig6}
\end{center}
\end{figure}
\section{conclusion}

With the help of a simple analytically solvable model of a spin star
system, we have considered layered environment model with one-qubit
in centr of layers and every layer is constructed by some spins. Of
course, we have assumed these layers have different radii which
means different coupling coefficients. By considering the Fig.(2)
and Fig.(4), and selecting equal degeneracy coefficients of both
layers and considering the influence of coupling coefficients in
decoherence-time control it can be deduced that the fluctuations in
$f_{z}$ is more intense than $f_{\pm}$. Meanwhile variations of
coupling coefficients, makes the fluctuations
\begin{figure}[htb]
\begin{center}
\includegraphics{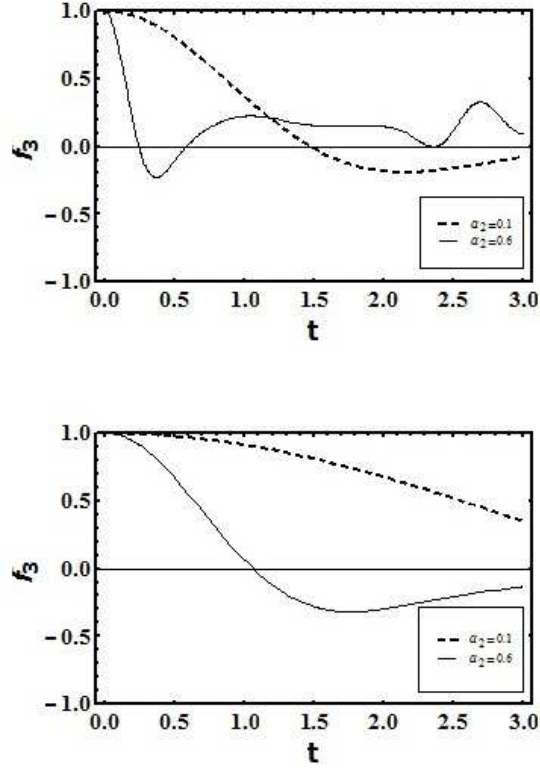}
      \vspace{10.5cm}
\caption{Comparison of limit $N\rightarrow\infty$ with limit
$N_{1}\rightarrow\infty$ or $N_{2}\rightarrow\infty$ for the changes
of correlation function $f_{3}$ [see Eqs.(39) and Eqs.(41)], with
$\alpha=0.1$ and $n=20$.} \label{fig7}
\end{center}
\end{figure}
of $f_{z}$ greater than $f_{\pm}$. In Fig.(4) one can see that the
increase of fluctuations can allow us to control, the decoherence.
but it's not good at $f_{\pm}$ because near zero, fluctuations are
gentle (this means that the amplitude of oscillation is shorter and
the course is slowly changing). Previous time of the system (at
short period of times) is longer than the decoherence time at
$f_{z}$ and is pulsed and is shorter at $f_{\pm}$ and continuous. As
a result, in general case for $f_{z}$ and $f_{\pm}$ to increasing
the power of coupling coefficient, yields greater control on
decoherence and increase of decoherence time. Also, with by
considering Fig.(3) and Fig.(5) and selecting inequal degeneracy
coefficients, we find that the same effects of selecting equal
degeneracy coefficients in Fig.(2) and Fig.(4) take place, but in
this part, fluctuations are more and the carve shifts towards the
vertical axis. It should again be emphasized that at $f_{z}$
transformation from coherence mode to decoherence mode and vice
versa is pulsed and at $f_{\pm}$ the decoherence is controlled
continuously. Then, we have considered the limit
$N\rightarrow\infty$ and for a special example (two layers with
different couple coefficients) showed that $N_{1}\rightarrow\infty$
with $N_{2}\rightarrow\infty$ have equal result. By considering
Fig.(6), the first chart is related to when $N_{1}\rightarrow\infty$
and $N_{2}\rightarrow\infty$, and the second chart is when
$N_{1}\rightarrow\infty$ or $N_{2}\rightarrow\infty$. In the first
case we can find that the fluctuations are increased by the increase
of coupling strength, this can also be seen in the second case too.
But the difference between first and second situation is in
decoherence-time control that in the first situation, this work is
well done because the curve in the first graph is distant from the
limit of zero and this means the increase in decoherence-time
control. Thus the best case is when $N_{1}\rightarrow\infty$ and
$N_{2}\rightarrow\infty$ occur simultaneously, which is more general
and real. It is much better because decoherence time is controled
better. In Fig.(7) we have the same conclusion. In Fig.(6) and
Fig.(7) we can see that we have more intensity in fluctuations of
$f_{z}$ than $f_{\pm}$, be side decoherence time at $f_{\pm}$ is
better than $f_{z}$ because at $f_{3}$ transformation of states from
coherence mode to decoherence mode and vice versa is pulsed.


\end{document}